\documentclass[twocolumn]{aastex63}
\usepackage{apjfonts}
\usepackage{graphicx}
\usepackage{xspace}

\begin{document}

\lefthead{{\em Chandra} Observations of IGR Sources}
\righthead{Tomsick et al.}

\def\lsim{\mathrel{\lower .85ex\hbox{\rlap{$\sim$}\raise
.95ex\hbox{$<$} }}}
\def\gsim{\mathrel{\lower .80ex\hbox{\rlap{$\sim$}\raise
.90ex\hbox{$>$} }}}

\title{{\em Chandra} Observations of High Energy X-ray Sources Discovered by {\em INTEGRAL}}

\author{John A. Tomsick}
\affiliation{Space Sciences Laboratory, 7 Gauss Way, University of California, Berkeley, CA 94720-7450, USA}

\author{Arash Bodaghee}
\affiliation{Georgia College \& State University, CBX 82, Milledgeville, GA 31061, USA}

\author{Sylvain Chaty}
\affiliation{AIM, CEA, CNRS, Universit\'{e} Paris-Saclay, Universit\'{e} Paris Diderot, Sorbonne Paris Cit\'{e}, F-91191 Gif-sur-Yvette, France}
\affiliation{APC, Universit\'{e} Paris Diderot, CNRS/IN2P3, CEA/IRFU, Observatoire de Paris, Sorbonne Paris Cit\'{e}, Paris 75205, France}

\author{Ma\"ica Clavel}
\affiliation{Univ. Grenoble Alpes, CNRS, IPAG, F-38000 Grenoble, France}

\author{Francesca M. Fornasini}
\affiliation{Harvard-Smithsonian Center for Astrophysics, 60 Garden Street, Cambridge, MA 02138, USA}

\author{Jeremy Hare},
\affiliation{Space Sciences Laboratory, 7 Gauss Way, University of California, Berkeley, CA 94720-7450, USA}
\affiliation{NASA Goddard Space Flight Center, Greenbelt, MD 20771, USA}
\affiliation{NASA Postdoctoral Program Fellow}

\author{Roman Krivonos}
\affiliation{Space Research Institute, Russian Academy of Sciences, Profsoyuznaya 84/32, 117997 Moscow, Russia}

\author{Farid Rahoui}
\affiliation{Independent}

\author{Jerome Rodriguez}
\affiliation{AIM, CEA, CNRS, Universit\'{e} Paris-Saclay, Universit\'{e} Paris Diderot, Sorbonne Paris Cit\'{e}, F-91191 Gif-sur-Yvette, France}

\begin{abstract}

The {\em International Gamma-Ray Astrophysics Laboratory (INTEGRAL)} satellite has detected in excess of 1000 sources in the $\sim$20--100\,keV band during its surveys of the sky over the past 17 years.  We obtained 5\,ks observations of 15 unclassified IGR sources with the {\em Chandra X-ray Observatory} in order to localize them, to identify optical/IR counterparts, to measure their soft X-ray spectra, and to classify them.  For 10 of the IGR sources, we detect {\em Chandra} sources that are likely (or in some cases certain) to be the counterparts.  IGR~J18007--4146 and IGR~J15038--6021 both have {\em Gaia} parallax distances, placing them at $2.5^{+0.5}_{-0.4}$ and $1.1^{+1.5}_{-0.4}$\,kpc, respectively.  We tentatively classify both of them as intermediate polar-type Cataclysmic Variables.  Also, IGR~J17508--3219 is likely to be a Galactic source, but it is unclear if it is a Dwarf Nova or another type of transient. For IGR~J17118--3155, we provide a {\em Chandra} localization, but it is unclear if the source is Galactic or extragalactic.  Based on either near-IR/IR colors or the presence of extended near-IR emission, we classify four sources as Active Galactic Nuclei (IGR~J16181--5407, IGR~J16246--4556, IGR~J17096--2527, and IGR~J19294+1327), and IGR~J20310+3835 and IGR~J15541--5613 are AGN candidates.  In addition, we identified an AGN in the {\em INTEGRAL} error circle of IGR~J16120--3543 that is a possible counterpart.

\end{abstract}


\section{Introduction}

Since 2002, the {\em INTEGRAL} satellite has been carrying out observations with its large field of view coded aperture mask instruments \citep{winkler03}.  In particular, as the exposure time in the $\sim$20-100\,keV band across the sky and especially in the Galactic plane has increased, more new or previously poorly studied ``{\em INTEGRAL} Gamma-Ray'' (IGR) sources have been found.  {\em INTEGRAL} has now detected in excess of 1000 sources \citep{bird16,krivonos17} with the majority being new sources.

The hard X-ray bandpass of {\em INTEGRAL} provides a way to select sources where extreme physics is occurring.  The high energy emission can be produced by accretion onto a compact object (magnetic white dwarf, neutron star, or black hole) or when particle acceleration leads to non-thermal emission.  The \cite{bird16} {\em INTEGRAL} source catalog includes 939 sources.  Over the whole sky, Active Galactic Nuclei (AGN) dominate and account for 369 of the sources.  The Galactic population is dominated by 129 low mass X-ray binaries (LMXBs), 116 high mass X-ray binaries (HMXBs), and 56 cataclysmic variables (CVs).  Pulsar wind nebulae (PWNe), supernova remnants (SNRs), and galaxy clusters are also among the IGR sources.  Thus, {\em INTEGRAL} is providing a much more complete picture of the populations of hard X-ray sources, and making discoveries about Galactic sources, including two new classes of HMXBs: obscured HMXBs \citep{mg03,walter06} and Supergiant Fast X-ray Transients \citep{negueruela06,sguera06,romano14}; larger numbers of hard X-ray emitting intermediate polars \citep[IPs;][]{bb07,landi09,tomsick16b}; and highly energetic PWNe \citep{tomsick12,pavan13}.

While {\em INTEGRAL} excels at detecting sources in the 20-100\,keV band, it only localizes the sources to $1^{\prime}$--$5^{\prime}$, which is not adequate for finding optical/IR counterparts.  Thus, in most cases, sources are not classified until higher angular resolution X-ray observations are obtained, providing an image to identify extended sources or to improve the source localization, allowing for multi-wavelength counterparts to be found.  Large X-ray follow-up programs include the use of the {\em Neil Gehrels Swift Observatory} X-ray Telescope \citep[{\em Swift}/XRT;][and references therein]{landi17} and the {\em Chandra X-ray Observatory} \citep[e.g.,][]{tomsick06,tomsick08,tomsick16}.  The X-ray localizations enable optical and near-IR spectroscopy \citep[e.g.,][]{chaty08,masetti13,coleiro13,fortin18}, at which point confident source classifications are nearly always obtained.

\subsection{Target Selection}

When the \cite{bird16} catalog was released, there were 219 unidentified sources in the catalog, and we selected sources to observe with {\em Chandra} based on the following criteria:

1. the selected sources have IGR names, indicating discovery in the hard X-rays by {\em INTEGRAL};

2. the nature is completely unknown for the selected sources (i.e., the type is listed as ``?'');

3. the sources are ``new'' in that they did not appear in the \cite{bird10} catalog, implying that there has been little opportunity for follow-up observations; and

4. they are within $15^{\circ}$ of the Galactic plane, increasing the chances that they are Galactic (we note that the highest Galactic latitude in our final list is $b = 11.35^{\circ}$).  

After applying these four criteria, 56 sources remained, but we reduced the list further based on available information about performed or planned observations with {\em Chandra}, the X-ray Multi-Mirror Mission ({\em XMM-Newton}), and {\em Swift}.  For example, when we selected targets, 20 of the 56 sources had planned {\em Swift} observations, and these were removed from our list (although we kept some sources that already had {\em Swift} observations after confirming that the {\em Swift} localization did not identify a unique optical or near-IR counterpart).  We also found that some of the sources show evidence for being variable or strongly variable in the hard X-rays based on the analysis of \cite{bird16}.  Strong variability (changes in {\em INTEGRAL} detection significance over time by $>$400\%) could indicate that the source is truly transient, which means that it could be too faint to be detected in a short {\em Chandra} observation, and so none of the sources in our final list are in the strongly variable category.  However, most Galactic sources (and certainly X-ray binaries) are variable, so we kept the sources in the variable category (changes in {\em INTEGRAL} detection significance over time by between 10\% and 400\%), and these are indicated in Table~\ref{tab:bird}.

After the above considerations, the list of 15 IGR sources in Table~\ref{tab:bird} remained. They were observed with {\em Chandra}, and we describe the observations in Section 2.  In Section 3, we describe the analysis of the data and present the results.  The primary goal is to obtain X-ray positions of the IGR sources with arcsecond accuracy, but this also includes assessing which is the most likely {\em Chandra} counterpart for each IGR source.  For this, we follow methods that are similar to those we have developed in earlier works \citep{tomsick09a,tomsick12a}. We discuss the results in Section 4 and summarize our conclusions in Section 5.

\section{Chandra Observations}

The list of IGR targets and information about the {\em Chandra} observations is provided in Table~\ref{tab:obs}.  For all 15 observations, the target was observed with the ACIS-I instrument \citep{garmire03}, which has a bandpass of 0.3--10\,keV.  IGR sources are typically relatively bright, with X-ray fluxes of $\sim$$10^{-12}$\,erg\,cm$^{-2}$\,s$^{-1}$.  At these flux levels, exposure times of $\sim$5\,ks are sufficient to detect tens or hundreds of counts, depending on spectral shape and variability.  Although some highly absorbed or highly variable IGR sources could have significantly lower X-ray flux levels, longer exposure times would not necessarily lead to the identification of the {\em Chandra} counterpart. With the 90\% confidence {\em INTEGRAL} error circles having radii between $2.31^{\prime}$ and $4.69^{\prime}$, we would expect the sky density of field sources to make it difficult to select the true counterpart if it had a significantly lower X-ray flux level than expected.

Table~\ref{tab:obs} also provides the pointing positions used for each IGR source.  In 11 cases, we simply used the {\em INTEGRAL} position reported in the \cite{bird16} catalog.  However, for four sources (IGR~J19294+1327, IGR~J16181--5407, IGR~J20310+3835, and IGR~J18007--4146), {\em Swift}/XRT observations identified a possible counterpart, and we used the XRT position for the {\em Chandra} pointing.  However, in each case, the full {\em INTEGRAL} error circle was included in the ACIS-I field of view, and in searching for the IGR source, we considered our search region to be the {\em INTEGRAL} error circle.

\section{Analysis and Results}

\subsection{Chandra Source Detection, Localization, and Photometry}

We reduced the data using the {\em Chandra} Interactive Analysis of Observations \citep[CIAO,][]{fruscione06} version 4.11 software along with the Calibration Database (CALDB) version 4.8.4.1.  We used {\ttfamily chandra\_repro} to reprocess the data, which results in an event list along with several other files with information about the spacecraft aspect, bad detector pixels, etc.  After reprocessing, the steps in our analysis are to search for sources on the four ACIS-I detector chips, cross-correlate with the {\em Gaia} Data Release 2 (DR2) optical source catalog \citep{gaia2,gaia_dr2} to register the images and reduce the contribution of the systematic pointing uncertainty to the source localization uncertainties, and perform aperture photometry to determine the number of counts for each detected source.  We generally follow the instructions in the CIAO science threads\footnote{See http://asc.harvard.edu/ciao/threads/index.html}.

For each of the 15 observations, we used {\ttfamily fluximage} to produce an exposure-corrected image in the 0.3--10\,keV energy band, and applied the {\ttfamily wavdetect} source detection algorithm to this image with wavelet scales of 1, 2, 4, 6, 8, 12, 16, 24, and 32 and the detection threshold set at a level to produce a list of sources for which only one spurious source is expected.  Table~\ref{tab:shifts} shows that between 6 and 28 {\em Chandra} sources were detected per ObsID.  We used the imaging program SAOImage ({\ttfamily ds9}) to search the {\em Gaia} DR2 catalog for optical sources in the ACIS-I field and cross-correlated the {\em Chandra} and {\em Gaia} source lists using CIAO's {\ttfamily wcs\_match}.  For 11 of the ObsIDs, between 3 and 7 matches were found, allowing for a position shift to be calculated to register the {\em Chandra} image to the {\em Gaia} reference frame.  The shifts are listed in Table~\ref{tab:shifts} along with the average residuals between the {\em Chandra} and {\em Gaia} positions, which are between $0.32^{\prime\prime}$ and $0.63^{\prime\prime}$. We take these values to be the systematic pointing uncertainty.  The four remaining ObsIDs have zero or one match, and we do not perform any position shifts for these.  Thus, the systematic pointing uncertainty in these four cases is $0.8^{\prime\prime}$ (90\% confidence)\footnote{See http://cxc.cfa.harvard.edu/cal/ASPECT/celmon/}.

We carried out the {\em Chandra} aperture photometry to determine the number of counts for all of the detected sources.  We made a point spread function (PSF) map using {\ttfamily mkpsfmap} for an energy of 2.3\,keV (the typical average photon energy for the full 0.3--10\,keV {\em Chandra} bandpass), and determined the 95\% encircled energy radius for each source.  After defining large (typically $\sim$15\,arcmin$^{2}$) source-free background regions for each observation, we used {\ttfamily dmextract} to extract background-subtracted counts in the 0.3--2\,keV, 2--10\,keV, and 0.3--10\,keV energy bands.  Based on our previous work \citep[e.g.,][]{tomsick12a}, for each source, we calculated the probability that the source would be detected in a search area with a radius of $\theta_{\rm search}$.  In cases where the source is within the 90\% confidence {\em INTEGRAL} error radius ($\theta_{INTEGRAL}$), $\theta_{\rm search} = \theta_{INTEGRAL}$.  If the source is outside the {\em INTEGRAL} error radius then $\theta_{\rm search}$ is equal to the angular distance from the best estimate of the {\em INTEGRAL} position.  

The predicted surface density of sources is the other factor that is important for determining the spurious source probability.  In \cite{tomsick12a}, we used
\begin{equation}
N(>F_{2-10~\rm keV}) = 9.2(F_{2-10~\rm keV}/10^{-13})^{-0.79}\,{\rm deg}^{-2}~~~,
\end{equation}
which is based on the $\log{N}$-$\log{S}$ curve from the {\em Advanced Satellite for Cosmology and Astrophysics} \citep[{\em ASCA};][]{sugizaki01}.  However, this work only included the Galactic sources, so we also consider a more recent determination of the $\log{N}$-$\log{S}$ in the Norma Region of the Galactic Plane
\begin{equation}
N(>F_{2-10~\rm keV}) = 36(F_{2-10~\rm keV}/10^{-13})^{-1.24}\,{\rm deg}^{-2}~~~,
\end{equation}
which also includes AGN \citep{fornasini14}.  In addition, the Norma Region is centered on the Galactic plane, covering --$0.4^{\circ} < b < 0.4^{\circ}$, likely representing an upper bound on the source density.  While we use these exact equations in Section 3.2 to determine the absolute spurious source probabilities for {\em Chandra} sources that are likely candidates to be associated with the IGR sources, we first determine the relative probabilities for all sources using
\begin{equation}
P_{\rm rel} = 1 - e^{-(\frac{C_{2-10\,\rm keV}}{C_{0}})^{-1.0}~\pi~\theta_{\rm search}^{2}}~~~,
\end{equation}
where $C_{2-10\,\rm keV}$ is the number of counts in the 2--10\,keV band, $C_{0}$ is an arbitrary normalization constant set so that the brightest sources have $P_{\rm rel}$ values near 1\%, and we use --1.0 as the slope of the $\log{N}$-$\log{S}$ since this is the midpoint between the \cite{sugizaki01} and \cite{fornasini14} values.  

\begin{figure}
\plotone{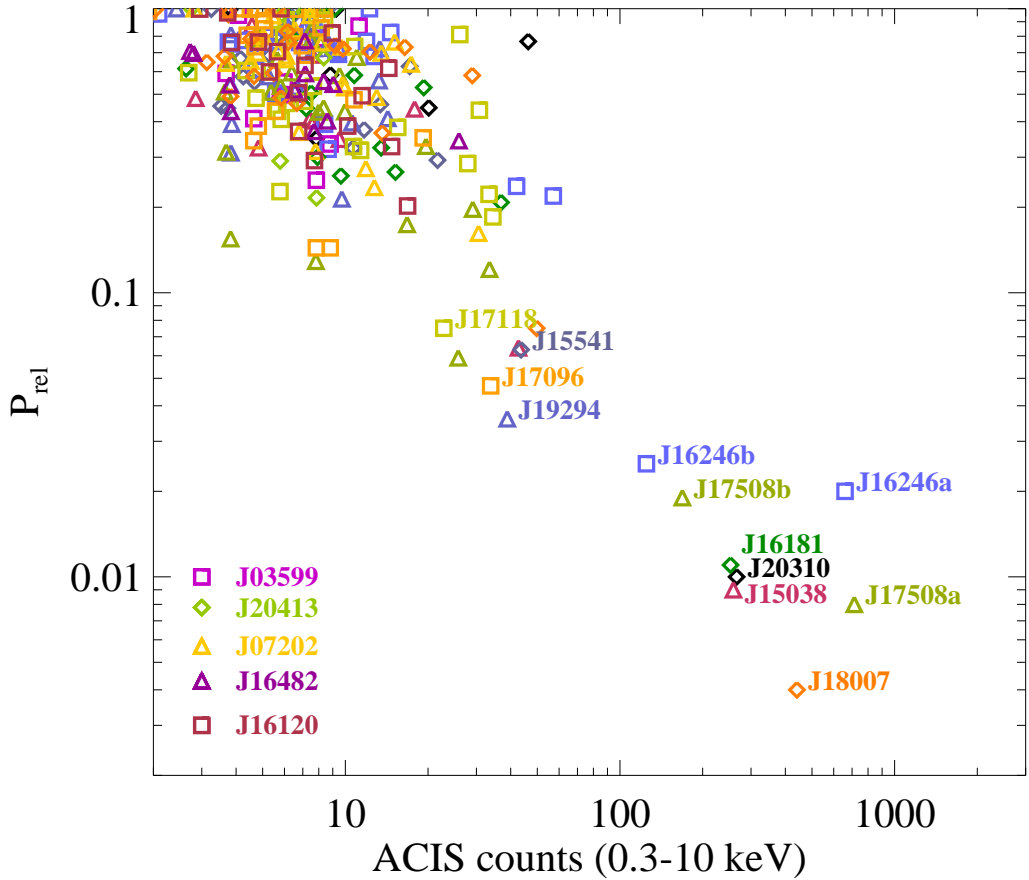}
\caption{The relative probability of a chance detection (see Equation 3) vs. the number of ACIS-I counts in the 0.3--10\,keV bandpass for all {\em Chandra} sources detected in the 15 observations of IGR source fields.  For the ten fields with candidate counterparts, the source that is least likely to be spurious is labeled with the IGR source name.  In two fields (J16246 and J17508), there are two sources that are potential counterparts.  The legend in the lower left corner of the plot only includes the five IGR source fields without likely counterparts.\label{fig:prel}}
\end{figure}

Using Equation 3, with $C_{0} = 140$, we calculate $P_{\rm rel}$ for all the sources detected in the 15 {\em Chandra} observations, and these are plotted in Figure~\ref{fig:prel}.  This parameter space leads to a large cluster of low count/high $P_{\rm rel}$ field sources and relatively clear separation between this cluster and the {\em Chandra} sources that are candidates to be counterparts to the IGR sources.  For ten of the IGR fields, the most likely counterpart is labeled in Figure~\ref{fig:prel}, and there are two fields with two potential counterparts. 

For these 12 sources, Table~\ref{tab:positions} gives the {\em Chandra} names and positions, the angular distance from the center of the {\em INTEGRAL} error circle ($\theta$), the number of ACIS counts in the 0.3--10\,keV band, and the hardness ratio.  The uncertainties in the positions include systematic and statistical contributions added in quadrature.  The determination of the systematic uncertainty is described above (see Table~\ref{tab:shifts}).  For the statistical contribution, we use Equation 13 from \cite{kim07a}, which uses the number of counts and the angular distance of the source from the {\em Chandra} aimpoint.  The information in Table~\ref{tab:positions} provides additional information about the likelihood that the {\em Chandra} sources are the true counterparts to the IGR sources.  In particular, J16246a is a questionable counterpart because it is one of the softer sources in the list, and it is well outside the {\em INTEGRAL} error circle ($\theta = 7.92^{\prime}$ compared to the error circle radius of $4.61^{\prime}$).  Also, J17508a is questionable since it is a very soft source, and it is outside the {\em INTEGRAL} error circle.

\subsection{Chandra and INTEGRAL Energy Spectra}

For the 12 candidate counterparts, we extracted {\em Chandra} energy spectra using {\ttfamily specextract}.  For the source extraction region, we used a circle with a radius corresponding to 95\% encircled energy, which we determined from the PSF map.  However, we also checked each source for photon pile-up by making an image for energies greater than 10\,keV.  For the three sources with evidence for pile-up (J16181, J18007, and J20310), we used an annular source extraction region with an inner radius of 1 pixel to cut out the core of the PSF.  We also extracted a background spectrum using the same region used for the photometry.  The spectra were rebinned to require a detection in each bin at the 3$\sigma$ level unless this resulted in fewer than five bins, in which case a 2$\sigma$ requirement was used.

Using XSPEC \citep{arnaud96}, we fit the {\em Chandra} spectra with an absorbed power-law model, and the parameters are reported in Table~\ref{tab:spectra_ch}.  We performed the fits by minimizing the C-statistic, and we give the $C$ values and number of degrees of freedom (dof) in Table~\ref{tab:spectra_ch}.  We use these values along with the variances in $C$, calculated according to equations 20-22 in \cite{kaastra17}, to quantify the quality of the fits.  The $P_{\rm reject}$ values in Table~\ref{tab:spectra_ch} indicate the probability that an absorbed power-law does not provide a good description of the spectrum.  The most significant deviations from an absorbed power-law are for J16246a, J18007, and J17508a, which have $P_{\rm reject}$ values of 99.7\%, 88\%, and 84\%, respectively.  While an absorbed power-law fit is not a formally acceptable model in these cases, the residuals do not show clear evidence for spectral features such as emission lines.  

The measured $N_{\rm H}$ values exceed the Galactic column density (see Table~\ref{tab:spectra_ch} for both) for J16181 ($N_{\rm H} = (9\pm 7)\times 10^{22}$\,cm$^{-2}$; 90\% confidence errors are given) and J19294 ($N_{\rm H} = (2.8^{+3.8}_{-2.4})\times 10^{23}$\,cm$^{-2}$).  Several other sources have upper limits on $N_{\rm H}$ that are higher than the Galactic $N_{\rm H}$ \citep{bb16}, but we note that there may be spatial variations in $N_{\rm H}$ on scales smaller than the angular resolution of the survey. The measured $\Gamma$ values show that most of the spectra are intrinsically hard.  Nine of the sources have photon indices with best fit values less than or equal to 1.5.  J17058a is the softest source with $\Gamma = 3.0\pm 0.2$.  While J17096 and J17508b have $\Gamma = 1.9\pm 1.3$ and $1.8\pm 0.5$, respectively, the errors are relatively large.  We use these fits to determine the fluxes in the 2--10\,keV band.  The ``unabsorbed'' fluxes are only corrected for the interstellar (Galactic) column density.  With these fluxes, we use
\begin{equation}
P = 1 - e^{-N(>F_{2-10~\rm keV})~\pi~\theta_{\rm search}^{2}}
\end{equation}
to assess the absolute probability of finding a source in the search region (defined by $\theta_{\rm search}$) as bright as the candidate sources.  We find $N(>F_{2-10~\rm keV})$ using the unabsorbed fluxes in Table~\ref{tab:spectra_ch} and Equations 1 and 2, which provides a range of probabilities depending on whether we use the \cite{sugizaki01} or the \cite{fornasini14} expression.  The range of probabilities for each source is given in Table~\ref{tab:spectra_ch}.  The least likely sources to be chance coincidences are J15038 (0.97--1.03\%), J18007 (1.3--1.7\%), and J20310 (1.5--1.6\%), and the sources that are most likely to be chance coincidences are J15541 (5.4--10.7\%) and J17096 (4.7--13.7\%).

While these are the formal probabilities, we can also obtain information by exploring how the {\em Chandra} spectra extrapolate into the {\em INTEGRAL} band.  To accomplish this comparison, we produced 2-bin {\em INTEGRAL} spectra using the fluxes in the \cite{bird16} catalog, and refit the {\em Chandra} spectra jointly with these two higher-energy points.  An important caveat is that the {\em INTEGRAL} fluxes are averages over 7--8 years (up to 2010) while the {\em Chandra} spectrum is from a single observation in 2017.  While the source flux (and possibly spectral shape) may be different for {\em Chandra} and {\em INTEGRAL}, we carry out fits without allowing for an offset between the two.  Thus, a mismatch between {\em Chandra} and {\em INTEGRAL} could be caused by variability or it could indicate that the {\em Chandra} counterpart is incorrect.  Also, for many of the {\em INTEGRAL} spectra, the flux is dropping between the 20--40 and 40--100\,keV energy bands, indicating that the spectrum must change slope between {\em Chandra} and {\em INTEGRAL}.  For this reason, we use a power-law with an exponential cutoff, and fit each spectrum with the XSPEC model {\ttfamily tbabs*cutoffpl}.

Figure~\ref{fig:spectra} shows the {\em Chandra}+{\em INTEGRAL} spectra, and the parameters are given in Table~\ref{tab:spectra}.  We also used XSPEC for these fits, but we minimized $\chi^{2}$ instead of $C$ because the {\em INTEGRAL} points are based on fluxes rather than Poissonian counts.  In nearly all cases, the {\ttfamily tbabs*cutoffpl} model provides a reasonably good description of the spectra.  One notable exception is J17508a, for which the power-law index is well-constrained by the {\em Chandra} spectrum to be soft, and the extrapolation into the {\em INTEGRAL} bandpass is below the {\em INTEGRAL} fluxes by orders of magnitude.  Although this might indicate that J17508b is the correct counterpart, the model does not fit the {\em Chandra}+{\em INTEGRAL} spectrum very well for J17508b either ($\chi^{2}/\nu = 43/14$), and we discuss the case of IGR~J17508--3219 further in Section~4.1.  J17096 is another case where the {\ttfamily tbabs*cutoffpl} model does not provide a very good fit, but it is clear that a better spectrum is required to understand the cause of the residuals.  It is possible that J17096 and J18007 both require an extra spectral component at low energies.

\subsection{Optical/IR Identifications}

We used the VizieR database to search for optical/IR counterparts to the 12 {\em Chandra} sources.  In seven cases, a {\em Gaia} DR2 optical source with a position consistent with the {\em Chandra} position was found, and these are listed in Table~\ref{tab:gaia}.  Parallax measurements are available for six of the sources.  They are negative in two cases, but for the four sources with positive parallaxes, the distance estimates from \cite{bj18} are $1.1^{+1.5}_{-0.4}$, $2.20^{+0.28}_{-0.22}$, $0.135^{+0.002}_{-0.001}$, and $2.5^{+0.5}_{-0.4}$\,kpc for J15038, J16246a, J17508a, and J18007, respectively.  Thus, all four sources are Galactic, but it is doubtful that J17508a\footnote{Catalog searches also uncovered that this source is the star HD 162186.} is the correct counterpart to the IGR source, and J16246a is a questionable counterpart.  However, IGR~J15038--6021 is a Galactic source at a distance of $1.1^{+1.5}_{-0.4}$\,kpc, and IGR~J18007--4146 is a Galactic source at a distance of $2.5^{+0.5}_{-0.4}$\,kpc.

Six of the 12 {\em Chandra} sources are present in the AllWISE IR catalog \citep{cutri14}, and the {\em Wide-field Infrared Survey Explorer (WISE)} magnitudes at 3.4$\mu$m ($W1$), 4.6$\mu$m ($W2$), 12$\mu$m ($W3$) and 22$\mu$m ($W4$) are given in Table~\ref{tab:wise}.  The {\em WISE} colors $W3$-$W2$ and $W2$-$W1$ have been used to identify AGN and blazars \citep[e.g.,][]{massaro12,secrest15,massaro16}, and we plot these colors for the six sources in Figure~\ref{fig:wise}.  It has already been reported by \cite{ursini18} that IGR~J16181--5407 (and the {\em Chandra} source J16181) is likely to be an AGN based on its {\em WISE} colors, and Figure~\ref{fig:wise} shows that the same criterion leads to IGR~J17096--2527 (J17096) being a likely AGN.  Figure~\ref{fig:wise} also shows that the two Galactic sources (based on {\em Gaia} as discussed above) are far away from the AGN region.  J16246b and J19294 are the other two sources in the AllWISE catalog, but we cannot conclude on their nature based on their location in the {\em WISE} color-color diagram.  However, we note that all four AGN candidates (J16181, J17096, J16246b, and J19294) are listed in AllWISE as having spatial profiles that are inconsistent with a point source in at least one photometric band, and the two Galactic sources (J16246a and J17508a) do not show evidence for extension (see Table~\ref{tab:wise}).

\begin{figure}
\plotone{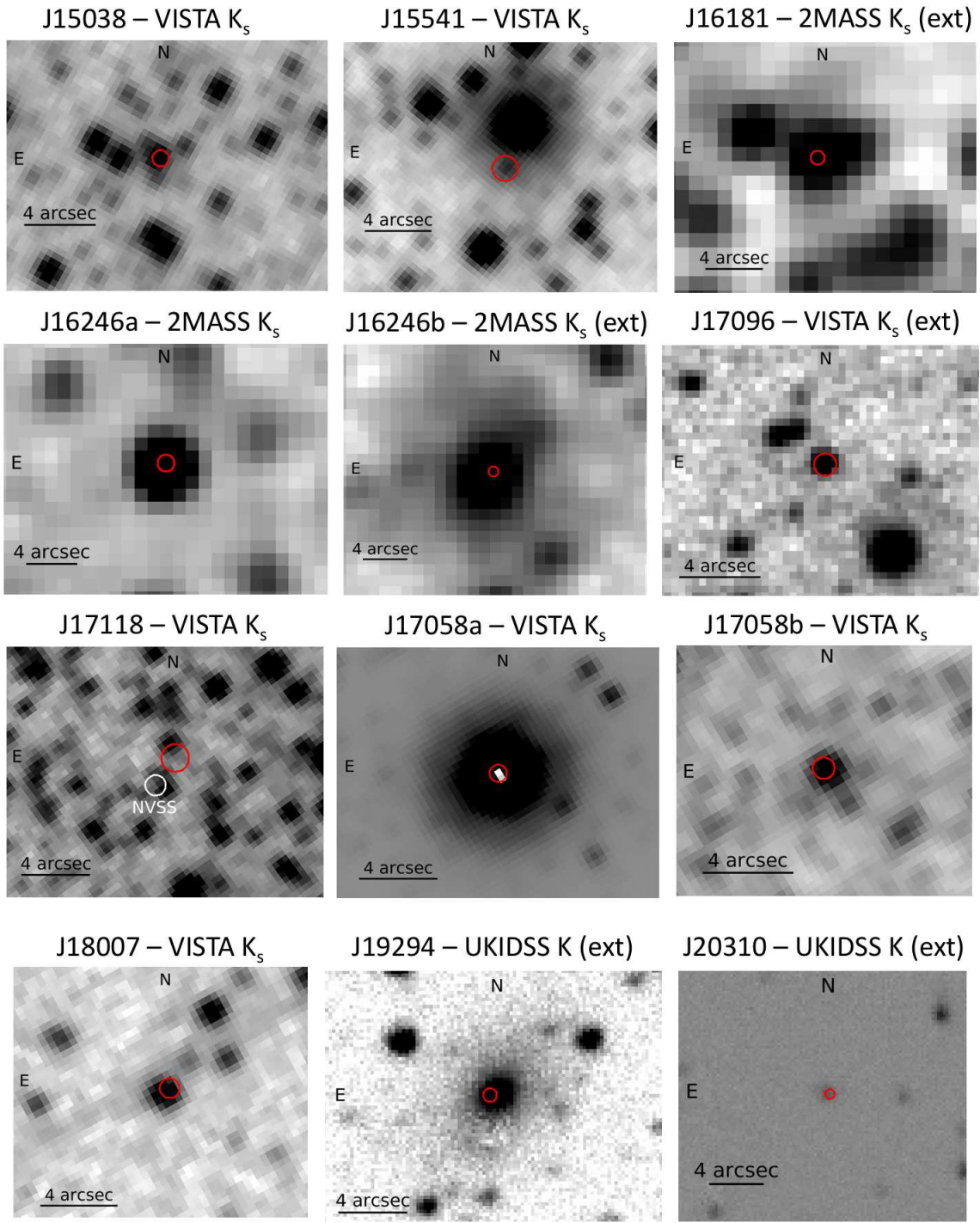}
\caption{$K$ or $K_{s}$ band images from the VISTA, 2MASS, and UKIDSS surveys for the 12 candidate counterparts.  The green circles indicate the {\em Chandra} positions (90\% confidence).  The sources that are extended (i.e., with spatial distributions that are not point-like; see Table~\ref{tab:nir}) are J16181, J16246b, J17096, J19294, and J20310 and are indicated with ``(ext)'' in the titles above the images.\label{fig:nir}}
\end{figure}

In addition to {\em Gaia} and {\em WISE}, we found many more optical/IR counterparts to the 12 {\em Chandra} sources in the VizieR database.  Here, we focus on the near-IR information from the Visible and Infrared Survey Telescope for Astronomy \citep[VISTA;][]{minniti10,mcmahon13,minniti17}, the UKIRT Infrared Deep Sky Survey \citep[UKIDSS;][]{lucas08}, and the 2 Micron All-Sky Survey \citep[2MASS;][]{cutri03}, providing the $J$, $H$, and $K$/$K_{s}$ magnitudes in Table~\ref{tab:nir} and the $K$/$K_{s}$ images in Figure~\ref{fig:nir}.  In the $K$/$K_{s}$ bands, the sources range in brightness with the faintest sources being J17118 ($K_{s} = 16.97\pm 0.20$) and J20310 ($K = 16.54\pm 0.04$) and the brightest sources being J17508a ($K_{s} = 7.45\pm 0.03$) and J16246a ($K_{s} = 11.49\pm 0.02$).  Although J15541 does not appear in the catalogs, Figure~\ref{fig:nir} shows that it has a faint $K_{s}$-band counterpart that is very close to a bright star.  In addition, the $K$-band images for J16246b, J19294, and probably J16181 confirm that these sources are extended, making it very likely that they are AGN.  It is unclear from the VISTA image whether J17096 is extended, and it is also classified as being a star in the VISTA catalog (see Table~\ref{tab:nir}); however, we still consider it to be likely that J17096 is an AGN based on the {\em WISE} information.  For J20310, the $K$-band image from UKIDSS may show some extension, and the source is classified as being a galaxy in the UKIDSS catalog.

In summary, the parallax measurements for J15038, J16246a, J17058a, and J18007 show that these four {\em Chandra} sources are Galactic, and the $K$/$K_{s}$ images (Figure~\ref{fig:nir}) are consistent with this in that they do not show evidence that these sources are extended.  Based on the sources being extended and/or the {\em WISE} colors, J16181, J16246b, J17096, and J19294 are AGN, and J20310 is an AGN candidate based on the UKIDSS information.  We discuss the nature of the remaining three sources (J15541, J17118, and J17058b) below.

\section{Discussion}

The main goal of this work is to determine the nature of the 15 IGR sources (see Table~\ref{tab:obs}) by determining the most likely {\em Chandra} counterparts and then using the {\em Chandra} localizations along with information in on-line catalogs and databases to classify the sources.  There are two IGR sources with two possible {\em Chandra} counterparts: IGR~J16246--4556 and IGR~J17508--3219; and we discuss those cases first.  Then, we consider the two sources definitively classified as being Galactic: IGR~J18007--4146 and IGR~J15038--6021.  After that, we discuss the unclassified sources with {\em Chandra} counterparts: IGR~J15541--5613 and IGR~J17118--3155 and then the cases without likely {\em Chandra} counterparts.

\subsection{IGR sources with 2 candidate Chandra counterparts}

\begin{figure}
\plotone{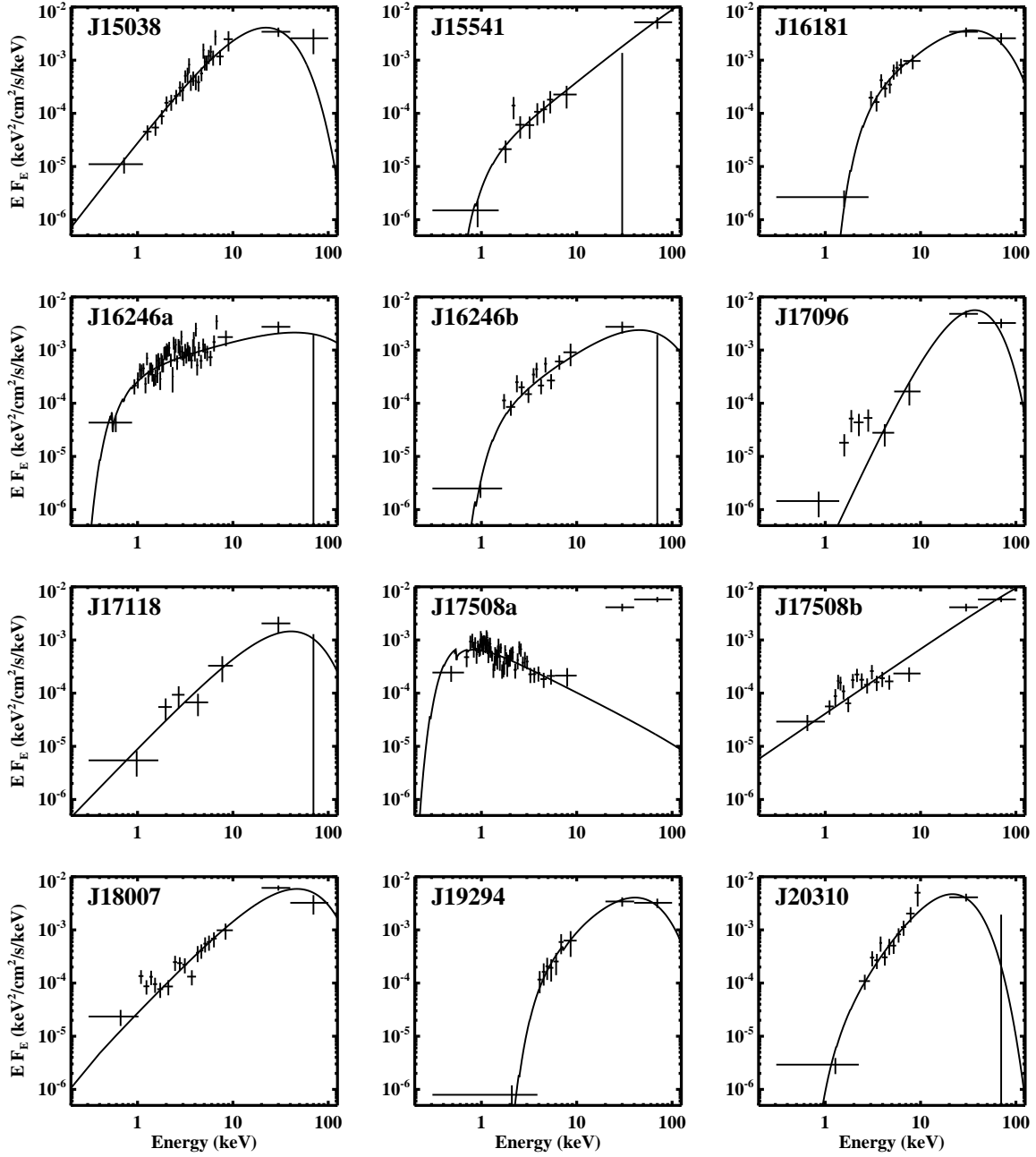}
\vspace{1.0cm}
\caption{{\em Chandra} and {\em INTEGRAL} energy spectra for the 12 candidate counterparts.  The two highest energy data points are derived from the fluxes reported in \cite{bird16}.  The best fit absorbed power-law with an exponential high-energy cutoff model ({\ttfamily tbabs*cutoffpl}) is shown.  We note that the {\em INTEGRAL} measurements were made during 2002--2010 while the {\em Chandra} spectra were obtained in 2017.  Thus, it is possible for source variability to cause poor fits in some cases.\label{fig:spectra}}
\end{figure}

For IGR~J16246--4556, the two candidate {\em Chandra} counterparts that we are considering are J16246a (3.0--3.5\%) and J16246b (2.9--3.5\%).  Both of them have hard spectra ($\Gamma = 1.5\pm 0.2$ and $1.3\pm 0.7$, respectively, from the {\em Chandra}-only fits) that are consistent with the {\em INTEGRAL} fluxes (see Figure~\ref{fig:spectra}), and their spurious probabilities are the same because J16246a is about three times brighter while J16246b is only 1.56$^{\prime}$ from the center of the {\em INTEGRAL} error circle, while $\theta=7.92^{\prime}$ for J16246a.  Given that the 90\% confidence {\em INTEGRAL} radius is $4.61^{\prime}$, an offset of $7.92^{\prime}$ corresponds to the 3$\sigma$ error radius; thus, it is unlikely that the source would be offset by this much, and we note that none of the other {\em Chandra} counterparts have values of $\theta$ that are more than $3.7^{\prime}$ (see Table~\ref{tab:positions}).  While J16246a may be an interesting source and worthy of attention, we identify J16246b with IGR~J16246--4556, making this IGR source an AGN.  The fact that J16246b is a galaxy is confirmed by the VizieR search, which shows a detection of the source in the Parkes HI zone of avoidance survey \citep{ss16}.  It is the galaxy HIZOA J1624-45B, which has a distance of 78.7\,Mpc.  Based on the {\em Chandra} and {\em INTEGRAL} spectrum, the unabsorbed 0.3--100\,keV flux is $7.7\times 10^{-12}$\,erg\,cm$^{-2}$\,s$^{-1}$, making the luminosity $5.7\times 10^{42}$\,erg\,s$^{-1}$.

IGR~J17508--3219 has the two candidate {\em Chandra} counterparts J17508a and J17508b.  Both are outside the $2.31^{\prime}$ (90\% confidence) {\em INTEGRAL} error circle with $\theta=3.21^{\prime}$ and $3.68^{\prime}$ for J17508a and J17508b, respectively.  As mentioned earlier, J17508a is a soft source ($\Gamma=3.0\pm 0.2$) and is inconsistent with the {\em INTEGRAL} fluxes (see Figure~\ref{fig:spectra}); thus, we rule out J17508a as the counterpart.  For J17508b, the VizieR search shows that it is coincident with the source OGLE-BLG-DN-0184, which is a dwarf nova (DN) non-magnetic CV with a few optical outbursts per year.  The typical outburst duration for OGLE-BLG-DN-0184 is 13.1 days where the optical brightness increases from $I = 17.5$ to 15.7 \citep{mroz15}.  Although 56 CVs appear in the \cite{bird16} catalog, nearly all of the CVs that have been classified are IPs or polars, which have highly magnetized white dwarfs.  However, some non-magnetic DNe can produce hard X-ray emission above at least 14\,keV \citep{mukai17}, and J17508b has a relatively hard spectrum with $\Gamma = 1.8\pm 0.5$.  Although we consider J17508b to be a possible counterpart, we also note that \cite{landi17} reported on a {\em Swift}/XRT observation where a third source (\#1 in the work of Landi) was detected within the {\em INTEGRAL} error circle with a 2--10\,keV flux of $8\times 10^{-13}$\,erg\,cm$^{-2}$\,s$^{-1}$ and $\Gamma = 0.6\pm 0.8$.  Landi source \#1 is not detected in the {\em Chandra} observation, and the upper limit on the 2--10\,keV flux is approximately an order of magnitude lower than the flux detected by {\em Swift}.  If Landi source \#1 is the true counterpart of IGR~J17508--3219, this level of variability is surprising given that it is noted to be a persistent source in \cite{bird16}.  Given that Landi source \#1 is a hard source within the {\em INTEGRAL} error circle, it must be considered as a strong candidate, but we also cannot rule out the possibility that J17508b also contributes to the flux seen by {\em INTEGRAL}.  

\subsection{Galactic Source IGR J18007--4146}

IGR~J18007--4146 is associated with an {\em XMM-Newton} slew source XMMSL1~J180042.8--414651, and the source was also detected by {\em Swift}/XRT \citep{landi17}.  Multiple optical/IR sources are consistent with the {\em XMM} and {\em Swift} localizations \citep{landi17}, but the {\em Chandra} position for CXOU~J180042.6--414650 that we report allows for a unique identification in multiple optical/IR surveys, including VISTA and {\em Gaia}.  It is the VISTA source VVV~J180042.71--414650.23, and the {\em Gaia} counterpart has a parallax distance measurement of $2.5^{+0.5}_{-0.4}$\,kpc.  The {\em Chandra} energy spectrum is a very hard power-law ($\Gamma<1$), and combining it with the {\em INTEGRAL} fluxes indicates that the spectrum turns over at around 25\,keV.  The 0.3--100\,keV flux is $(1.67\pm 0.35)\times 10^{-11}$\,erg\,cm$^{-2}$\,s$^{-1}$, and this corresponds to a luminosity of $(1.2\pm 0.5)\times 10^{34}$\,erg\,s$^{-1}$.  Known Galactic source types that may match these properties are CV/IPs or accreting pulsars, which are most often found in HMXBs.

The origin of the optical/IR emission is unclear.  If the source is a CV/IP, the emission would be a combination of an accretion disk and a star.  Most CV/IPs have late-type stars, and the accretion disk dominates.  If the source is an HMXB, then the high-mass star (O- or B-type) would dominate.  In either case, the emission is thermal, and we can estimate the temperature using the near-IR colors.  The X-ray spectrum does not show evidence for any absorption, with a 90\% confidence upper limit of $N_{\rm H} < 1.2\times 10^{21}$\,cm$^{-2}$, corresponding to $A_{V} < 0.54$ \citep{go09}.  Using the VISTA near-IR magnitudes and the \cite{ccm89} extinction law, the dereddened magnitudes are $J = 15.45$--15.60 and $H = 15.34$--15.24.  Thus, $J$--$H$ is between 0.21 and 0.26, which corresponds to a temperature of $\sim$6500\,K.  If this was a stellar temperature, it would correspond to an F5V spectral type, which has $M_{J} = 2.7$.  For IGR~J18007--4146, if the extinction is zero, then $M_{J} = 3.6\pm 0.4$, and if the extinction is at the maximum value ($A_{J} = 0.15$), it would be $M_{J} = 3.5\pm 0.4$, and these values correspond to a spectral type of G2V.  To accomodate a F5V star, the source distance would need to be $\sim$3.5\,kpc, which is higher than the \cite{bj18} value at the 2$\sigma$ level.  Regardless of whether the distance could be as high as 3.5\,kpc, these calculations show that IGR~J18007--4146 does not harbor a high-mass O or B type star.  In fact, if there is a significant contribution to the optical/IR emission from an accretion disk, it is possible that the source is a binary with a late-type star.  Given the possible classifications mentioned above (a CV/IP or an HMXB), a CV/IP is strongly favored.  However, follow-up optical spectroscopy is needed for confirmation.

\subsection{Galactic Source IGR J15038--6021}

IGR~J15038--6021 was also previously reported as a {\em Swift}/XRT source by \cite{landi17}, but the X-ray localization was not adequate to identify a unique optical/IR counterpart.  The detection of this source by {\em Chandra} as CXOU~J150415.7--602123 provides the unique optical/IR identification with the VISTA source VVV~150415.72--602122.87 as well as a {\em Gaia} source, which provides a parallax distance of $1.1^{+1.5}_{-0.4}$\,kpc.  The X-ray energy spectrum is slightly harder than IGR~J18007--4146, and the 0.3--100\,keV flux is $(1.11\pm 0.28)\times 10^{-11}$\,erg\,cm$^{-2}$\,s$^{-1}$, corresponding to a luminosity of $(1.6^{+4.4}_{-1.2})\times 10^{33}$\,erg\,s$^{-1}$.  This suggests the same possibilities for classifications mentioned above: HMXB or CV/IP.

We carry out a similar calculation as for IGR~J18007--4146 using the $J$ and $H$ magnitudes along with the {\em Gaia} distance.  Even though IGR~J15038--6021 is closer than IGR~J18007--4146, it is in the Galactic plane ($b$ = --$1.57^{\circ}$ compared to $b$ = --$9.12^{\circ}$ for IGR~J18007--4146), and although the best fit value for $N_{\rm H}$ is zero for both sources, the upper limit for IGR~J15038--6021 is higher ($N_{\rm H} < 1.0\times 10^{22}$\,cm$^{-2}$).  This limit on the column density corresponds to $A_{V}<4.52$ \citep{go09}, and we calculate dereddened magnitudes of $J = 14.86$--16.13 and $H = 14.62$--15.48, giving a range of $J$--$H$ values between 0.24 and 0.65 and a range of temperatures between 4000 and 6500\,K.  This allows for a range of stellar spectral types between K5V and F5V.  

Turning to the measurement of the absolute magnitude, assuming the {\em Gaia} distance ($1.1^{+1.5}_{-0.4}$\,kpc), if there is no extinction, then $M_{J} = 5.9^{+0.8}_{-3.0}$, and if the extinction is maximal ($A_{J} = 1.27$), then $M_{J} = 4.7^{+0.8}_{-3.0}$.  If the errors are included, these absolute magnitudes correspond to spectral types between K5V and A5V.  Although this is a large range, it does not cover O and B stars.  However, in this case, we must treat the {\em Gaia} distance with some caution because the {\em Gaia} DR2 catalog indicates that the astrometric noise for this source is significant.  Thus, we performed an additional calculation and found that the distance would need to be 10\,kpc to move the absolute magnitude of IGR~J15038--6021 into the late B-type range.  Thus, based on both the temperatures derived above and the fact that it is very unlikely for the true distance to be larger than the {\em Gaia} measurement by 6$\sigma$, we conclude that IGR~J15038--6021 is likely to be a CV/IP.

\subsection{Unclassified Sources with Chandra Counterparts}

Although our candidate {\em Chandra} counterpart to IGR~J15541--5613, CXOU~J155413.0--560932, has a relatively high spurious probability of 5.4--10.7\%, Figure~\ref{fig:spectra} shows that the spectrum rises throughout the {\em Chandra} band with $\Gamma = 1.1\pm 1.1$, and the source is also unusual for a Galactic source in having a higher flux in the 40--100\,keV band than in the 20--40\,keV band.  This suggests that IGR~J15541--5613 is more likely to be an AGN than a Galactic source.  It is conceivable that the high point in the {\em Chandra} spectrum at $\sim$2\,keV is a redshifted iron line, but the statistical significance of the putative line is low.  Also, the implied redshift is $z\sim 2$, and this would indicate a very luminous and rare AGN.  Figure~\ref{fig:nir} shows that, with the bright nearby source, obtaining an optical or near-IR spectrum of IGR~J15541--5613 will be challenging, but it may be possible.

For our candidate {\em Chandra} counterpart to IGR~J17118--3155, CXOU~J171135.8--315504, a search of the SIMBAD database shows a possible Very Large Array radio counterpart, NVSS~J171135--315506.  This radio source is quite bright, and is seen in several radio surveys \citep{intema17,douglas96,murphy07,condon98}: the Giant Metrewave Radio Telescope/GMRT (1496 mJy at 150 MHz); the Texas Survey/TXS (606 mJy at 365 MHz); the Molonglo Galactic Plane Survey/MGPS-2 (225 mJy at 843 MHz); and the NRAO VLA Sky Survey/NVSS (165 mJy at 1420 MHz). However, considering the best estimates for the radio positions, only the TXS position is within the {\em Chandra} error circle with the other sources being $2.3^{\prime\prime} \pm 2.0^{\prime\prime}$ (GMRT), $1.5^{\prime\prime} \pm 1.6^{\prime\prime}$ (MGPS-2), and $2.0^{\prime\prime} \pm 0.6^{\prime\prime}$ (NVSS) away from the center of the {\em Chandra} error circle.  The NVSS error circle is shown on the VISTA $K_{s}$ image (Figure~\ref{fig:nir}), and it is clear that the error region does not overlap with {\em Chandra}.  We also plotted the NVSS error circle on the {\em Chandra}/ACIS image, and none of the 23 ACIS counts fall within the NVSS error circle.  We repeated this analysis without the $0.3^{\prime\prime}$ position registration shift (see Table~\ref{tab:shifts}), and there are still no ACIS counts in the NVSS error circle.  While it would be a somewhat surprising coincidence for these two relatively unusual sources to be as close to each other as they are and not be associated, our analysis does not support an association between the two.

\subsection{IGR Sources without Likely Chandra Counterparts}

Although {\em Chandra} sources were detected in the {\em INTEGRAL} error circles for all 15 IGR sources included in this work, there were five cases (IGR~J03599+5043, IGR~J07202+0009, IGR~J16482--2959, IGR~J16120--3543, and IGR~J20413+3210) where the {\em Chandra} sources all had relatively high probability of being chance detections of field sources.  The {\em Chandra} sources in {\em INTEGRAL} error circles of these five IGR sources with the lowest values of $P_{\rm rel}$ are 24.9\%, 27.9\%, 37.0\%, 20.2\%, and 21.6\%, respectively.  We searched the VizieR database to check whether any of the {\em Chandra} sources have classified counterpart, and the source with $P_{\rm rel} = 20.2$\% (CXOU J161147.0--354634)\footnote{The {\em Chandra} position is R.A. = 16h11m47.04s, Decl. = --35$^{\circ}$46$^{\prime}$34.9$^{\prime\prime}$, equinox 2000.0 with a 90\% confidence position uncertainty of $0.90^{\prime\prime}$.} is a {\em WISE} source (AllWISE J161147.06--354635.0) that has been classified as an AGN based on its {\em WISE} colors \citep{secrest15}.  The 2--10\,keV X-ray flux is $(9.9^{+6.5}_{-7.1})\times 10^{-14}$\,erg\,cm$^{-2}$\,s$^{-1}$.  While this may still be a spurious identification, we consider the CXOU/AllWISE source to be a candidate counterpart.

Within the {\em INTEGRAL} error circles of the other four IGR sources (IGR~J03599+5043, IGR~J07202+0009, IGR~J16482--2959, and IGR~J20413+3210), the {\em Chandra} sources with the lowest $P_{\rm rel}$ values have between 5 and 7 counts in the 2--10\,keV band.  Taking 7 counts (0.0014\,c/s) as the upper limit, we can use PIMMS\footnote{See https://asc.harvard.edu/toolkit/pimms.jsp} to determine the approximate flux upper limit.  If we assume $\Gamma>0$ and $N_{\rm H} = 10^{22}$\,cm$^{-2}$, we find a flux limit of $<$$7\times 10^{-14}$\,erg\,cm$^{-2}$\,s$^{-1}$ in the 2--10\,keV energy band.  There are several possible interpretations for these IGR sources.  When similar results have previously been obtained for other IGR sources, \cite{tomsick16} consider that those IGR sources may be variable or may have hard spectra (either due to high column densities or hard power-law slopes).  It is notable that IGR~J03599+5043, IGR~J07202+0009, IGR~J16482--2959, and IGR~J20413+3210 are marked as being variable in \cite{bird16} (see Table~\ref{tab:bird}).  Another possibility is that these IGR sources are spurious, but the {\em INTEGRAL} signal-to-noise values are between 5.9 and 7.5 (Table~\ref{tab:bird}).  

\section{Summary and Conclusions}

\begin{figure}
\plotone{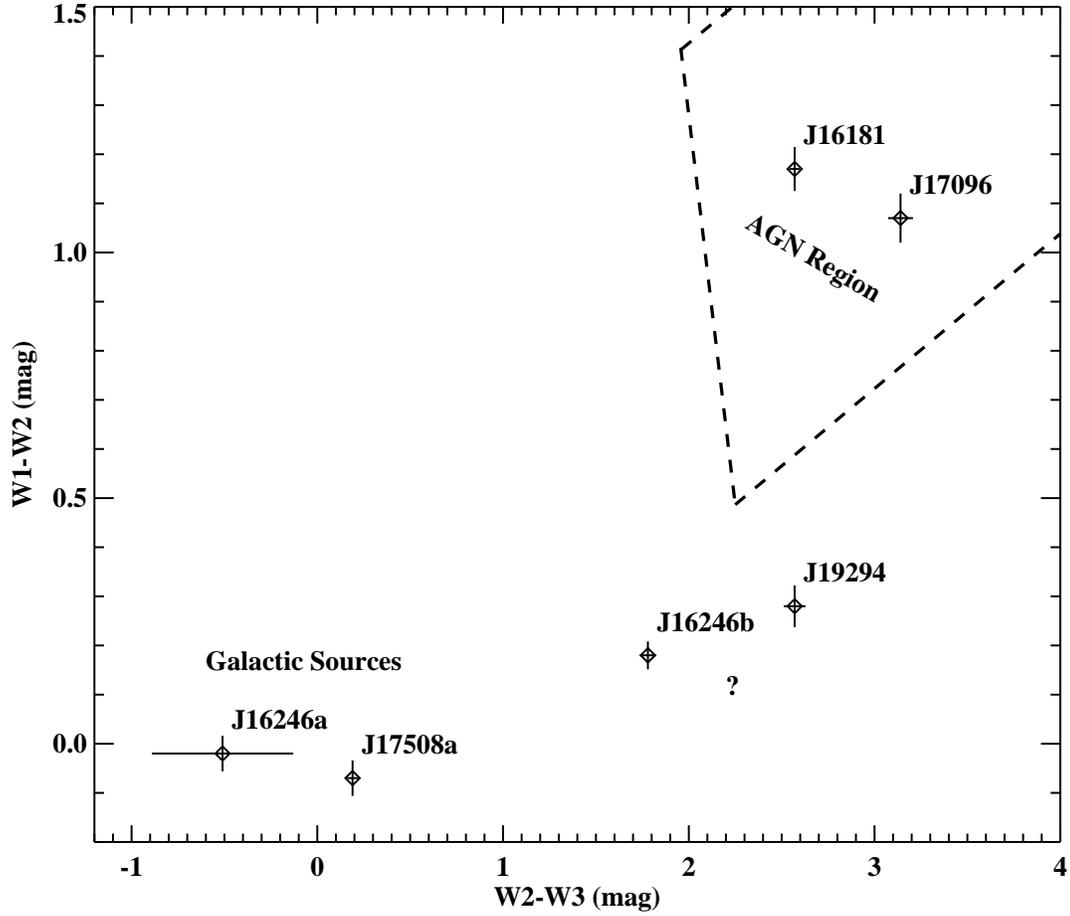}
\caption{Near-IR colors measured for six of the IGR sources by {\em WISE}.  $W1$, $W2$, and $W3$ are 3.4\,$\mu$m, 4.6\,$\mu$m, and 12\,$\mu$m, respectively.  Based on {\em Gaia} distances, J16246a and J17508a are Galactic sources (this work).  J16181 and J17096 are in a region commonly populated by AGN \citep{secrest15,ursini18}.  The region originally defined by \cite{mateos12} is indicated by the dashed lines.\label{fig:wise}}
\end{figure}

Our final results for each of the 15 IGR sources are summarized in Table~\ref{tab:summary}, including the name of the {\em Chandra} counterpart or upper limit on the X-ray flux, our conclusion about the source type, and the main evidence we use to come to our conclusion.  IGR~J15038--6021 and IGR~J18007--4146 are strong CV/IP candidates, and this could be confirmed in the future with optical or near-IR spectroscopy.  We definitively classify four IGR sources as AGN and three as AGN candidates.  This relatively high fraction of AGN may not be too surprising as the new IGR sources found recently are, on-average, fainter than the IGR sources found earlier in the mission.  We also note that the plot of near-IR colors measured by {\em WISE} (Figure~\ref{fig:wise}), which has been used to identify AGN, may be a useful tool for identifying Galactic sources as well.  Finally, although we detected likely {\em Chandra} counterparts to IGR~J17508--3219 and IGR~J17118--3155, the classifications are unclear.  Additional X-ray observations of IGR~J17508--3219 may be useful to determine if the true counterpart is CXOU~J175108.7--322122 or the \cite{landi17} {\em Swift} source.  For IGR~J17118--3155, the uncertainty on the X-ray position could be reduced with a longer {\em Chandra} observation with the source on-axis.

\clearpage

\acknowledgments

We thank the referee for helpful comments that improved the manuscript.  JAT and JH acknowledge partial support from the National Aeronautics and Space Administration (NASA) through {\em Chandra} Award Number GO7-18030X issued by the {\em Chandra} X-ray Observatory Center, which is operated by the Smithsonian Astrophysical Observatory under NASA contract NAS8-03060.  JH also acknowledges support from an appointment to the NASA Postdoctoral Program at the Goddard Space Flight Center, administered by the USRA through a contract with NASA.  SC is grateful to the Centre National d'Etudes Spatiales (CNES) for the funding of MINE (Multi-wavelength INTEGRAL Network). MC and JR acknowledge financial support from CNES.  RK acknowledges support from the Russian Science Foundation (grant 19-12-00396). This research has made use of the VizieR catalog access tool and the SIMBAD database, which are both operated at CDS, Strasbourg, France.  This work has made use of data from the European Space Agency (ESA) mission {\it Gaia} (\url{https://www.cosmos.esa.int/gaia}), processed by the {\it Gaia} Data Processing and Analysis Consortium (DPAC, \url{https://www.cosmos.esa.int/web/gaia/dpac/consortium}). Funding for the DPAC has been provided by national institutions, in particular the institutions participating in the {\it Gaia} Multilateral Agreement. This publication makes use of data products from the Wide-field Infrared Survey Explorer, which is a joint project of the University of California, Los Angeles, and the Jet Propulsion Laboratory/California Institute of Technology, funded by NASA.

\smallskip
\facilities{CXO, INTEGRAL, Gaia, WISE}

\smallskip
\software{CIAO (Fruscione et al. 2006), XSPEC (Arnaud 1996)}
  
\clearpage




\begin{table}
\caption{Source information from the 2016 {\em INTEGRAL} catalog\label{tab:bird}}
\begin{minipage}{\linewidth}
\footnotesize
\begin{tabular}{cccccccccc} \hline \hline
IGR Name & $l$\footnote{Galactic longitude converted from {\em INTEGRAL} position.}   & $b$\footnote{Galactic latitude converted from {\em INTEGRAL} position.}   &  RA\footnote{Source position measured by {\em INTEGRAL} and reported in \cite{bird16}.} & Dec$^{c}$  & Uncertainty\footnote{90\% confidence {\em INTEGRAL} error radius.} & Flux\footnote{The flux measured by {\em INTEGRAL} in units of mCrab.}         & Flux$^{e}$          & Variability\footnote{As described in \cite{bird16}, ``Y'' indicates that the source's detection significance in the {\em INTEGRAL} data is variable at a level of between 10\% and 400\%.} & Significance\footnote{The significance of the {\em INTEGRAL} detection in terms of signal-to-noise.}\\ 
         & (deg) & (deg) &  (deg)   & (deg)    & (arcmin)    & (20--40\,keV) & (40--100\,keV) &             &             \\ \hline\hline
J20310+3835  & 77.77  & --0.49 & 307.755 &  +38.576 & 4.54   & $0.6\pm 0.1$ & $<$0.3        & ---         & 5.5\\
J15038--6021 & 318.60 & --1.57 & 225.941 & --60.357 & 3.95   & $0.5\pm 0.1$ & $0.4\pm 0.2$  & ---         & 6.4\\
J03599+5043  & 150.58 & --1.74 &  59.973 &  +50.728 & 4.01   & $0.7\pm 0.2$ & $1.2\pm 0.3$  & Y           & 6.3\\
J15541--5613 & 326.38 & --1.90 & 238.527 & --56.216 & 4.32   & $<$0.2       & $0.8\pm 0.2$  & Y           & 5.8\\
J19294+1327  & 49.22  & --2.12 & 292.374 &  +13.451 & 3.43   & $0.5\pm 0.1$ & $0.5\pm 0.1$  & ---         & 7.5\\
J16246--4556 & 336.83 &  +2.46 & 246.091 & --45.923 & 4.61   & $0.4\pm 0.1$ & $<$0.3        & Y           & 5.4\\
J16181--5407 & 330.35 & --2.62 & 244.533 & --54.103 & 4.61   & $0.5\pm 0.1$ & $0.4\pm 0.1$  & ---         & 5.4\\
J17508--3219 & 357.68 & --2.72 & 267.721 & --32.330 & 2.31   & $0.6\pm 0.1$ & $0.9\pm 0.1$  & ---         & 11.6\\
J17118--3155 & 353.48 &  +4.41 & 257.959 & --31.927 & 2.84   & $0.3\pm 0.1$ & $<$0.2        & Y           & 9.2\\
J20413+3210  & 73.96  & --6.02 & 310.384 &  +32.219 & 3.42   & $<$0.3       & $<$0.4        & Y           & 7.5\\
J07202+0009  & 216.12 &  +6.47 & 110.063 &   +0.127 & 4.25   & $<$0.4       & $<$0.7        & Y           & 5.9\\
J17096--2527 & 358.47 &  +8.56 & 257.432 & --25.470 & 2.73   & $0.8\pm 0.1$ & $0.8\pm 0.1$  & ---         & 9.6\\
J18007--4146 & 350.38 & --9.12 & 270.202 & --41.802 & 3.19   & $0.9\pm 0.1$ & $0.5\pm 0.2$  & ---         & 8.1\\
J16482--2959 & 351.94 &  +9.52 & 252.145 & --30.019 & 4.25   & $0.3\pm 0.1$ & $0.6\pm 0.2$  & Y           & 5.9\\
J16120--3543 & 342.36 & +11.35 & 242.974 & --35.754 & 4.69   & $0.4\pm 0.2$ & $0.6\pm 0.3$  & Y           & 5.3\\ \hline
\end{tabular}
\end{minipage}
\end{table}

\begin{table}
\caption{{\em Chandra} Observations\label{tab:obs}}
\begin{minipage}{\linewidth}
\footnotesize
\begin{tabular}{cclccccc} \hline \hline
IGR Name & ObsID & Start Time (UT) & Exposure &  RA     & Dec      & Reference\\ 
         &       &                 & Time (s) &  (deg)  & (deg)    &          \\ \hline\hline
J20310+3835  & 18969 & 2017 Feb 26, 19.2\,h & 4880 & 307.73052 & +38.56095 & our analysis\footnote{We determined this position via analysis of {\em Swift}/XRT data.  The XRT position is now published in \cite{landi17}.}\\
J15038--6021 & 18970 & 2017 Apr 26, 13.5\,h & 4899 & 225.941 & --60.357 & \cite{bird16}\\
J03599+5043  & 18971 & 2017 Mar 20,  1.8\,h & 5056 &  59.973 &  +50.728 & \cite{bird16}\\
J15541--5613 & 18972 & 2017 May 22,  5.3\,h & 4995 & 238.527 & --56.216 & \cite{bird16}\\
J19294+1327  & 18973 & 2017 Feb 24,  5.9\,h & 4880 & 292.37417 & +13.45151 & \cite{pavan11}\\
J16246--4556 & 18974 & 2017 May 3,  12.6\,h & 4886 & 246.091 & --45.923 & \cite{bird16}\\
J16181--5407 & 18975 & 2017 May 20, 14.1\,h & 5019 & 244.53342 & --54.10272 & \cite{landi4233}\\
J17508--3219 & 18976 & 2017 May 8,  22.6\,h & 5013 & 267.721 & --32.330 & \cite{bird16}\\
J17118--3155 & 18977 & 2017 Jan 27, 16.8\,h & 4883 & 257.959 & --31.927 & \cite{bird16}\\
J20413+3210  & 18978 & 2017 Feb 24,  7.8\,h & 4730 & 310.384 &  +32.219 & \cite{bird16}\\
J07202+0009  & 18979 & 2017 Sep  9, 18.1\,h & 4899 & 110.063 &   +0.127 & \cite{bird16}\\
J17096--2527 & 18980 & 2017 Jan 25, 11.0\,h & 4845 & 257.432 & --25.470 & \cite{bird16}\\
J18007--4146 & 18981 & 2017 Aug 22,  5.2\,h & 4994 & 270.17792 & --41.78028 & our analysis$^{a}$\\
J16482--2959 & 18982 & 2017 Jun 20, 15.6\,h & 4979 & 252.145 & --30.019 & \cite{bird16}\\
J16120--3543 & 18983 & 2017 Mar 31, 16.8\,h & 4880 & 242.974 & --35.754 & \cite{bird16}\\ \hline
\end{tabular}
\end{minipage}
\end{table}

\begin{table}
\caption{Source Detection and Shifts Based on {\em Gaia} DR2 Matches\label{tab:shifts}}
\begin{minipage}{\linewidth}
\footnotesize
\begin{tabular}{ccccccc} \hline \hline
IGR Name & ObsID & $N_{Chandra}$\footnote{The number of {\em Chandra} sources detected on the four ACIS-I detector chips.} & $N_{\rm matches}$\footnote{The number of matches between the {\em Chandra} detections and the {\em Gaia} DR2 catalog.} & $x_{\rm shift}$\footnote{The shifts in the x and y detector coordinate directions in pixels.  The conversion is 1 pixel = $0.492^{\prime\prime}$.} & $y_{\rm shift}$$^{c}$ & Residual\footnote{Average residual (in arcseconds) between the {\em Chandra} and {Gaia} DR2 sources.}\\ \hline\hline
J20310+3835  & 18969 & 11 & 5 & --0.46 &   0.75 & 0.34\\
J15038--6021 & 18970 & 12 & 3 &   0.59 &   0.21 & 0.42\\
J03599+5043  & 18971 & 10 & 0 &    --- &    --- & ---\\
J15541--5613 & 18972 & 21 & 5 & --0.11 & --0.20 & 0.63\\
J19294+1327  & 18973 & 17 & 3 & --1.20 &   0.77 & 0.32\\
J16246--4556 & 18974 & 27 & 5 &   0.09 & --0.41 & 0.43\\
J16181--5407 & 18975 & 14 & 4 & --0.02 & --0.41 & 0.49\\
J17508--3219 & 18976 & 28 & 7 & --0.49 &   0.18 & 0.41\\
J17118--3155 & 18977 & 25 & 4 &   0.54 & --0.21 & 0.62\\
J20413+3210  & 18978 &  6 & 0 &    --- &    --- & ---\\
J07202+0009  & 18979 & 26 & 3 & --0.70 &   0.10 & 0.35\\
J17096--2527 & 18980 & 23 & 7 &   0.20 &   0.69 & 0.38\\
J18007--4146 & 18981 & 22 & 4 &   0.65 & --0.45 & 0.50\\
J16482--2959 & 18982 & 13 & 0 &    --- &    --- & ---\\
J16120--3543 & 18983 & 17 & 1 &    --- &    --- & ---\\ \hline
\end{tabular}
\end{minipage}
\end{table}

\begin{table}
\caption{{\em Chandra} Candidate Counterparts to IGR Sources\label{tab:positions}}
\begin{minipage}{\linewidth}
\footnotesize
\begin{tabular}{llcccccc} \hline \hline
IGR Name & CXOU Name & {\em Chandra} R.A. & {\em Chandra} Decl. & Uncertainty\footnote{90\% confidence.} & $\theta$\footnote{The angular distance between the center of the {\em INTEGRAL} error circle and the source.}/$\theta_{INTEGRAL}$\footnote{The size of the 90\% confidence {\em INTEGRAL} error radius given in \cite{bird16}.} & ACIS & Hardness\footnote{The hardness is given by $(C_{2}-C_{1})/(C_{2}+C_{1})$, where $C_{2}$ is the number of counts in the 2--10 keV band and $C_{1}$ is the number of counts in the 0.3--2 keV band.}\\
 & & (J2000) & (J2000) & (arcseconds) & (arcminutes) & Counts\footnote{The number of counts, after background subtraction, measured by {\em Chandra}/ACIS-I in the 0.3--10\,keV band.} & \\ \hline \hline
J15038 & J150415.7--602123 & 15h04m15.70s & --60$^{\circ}$21$^{\prime}$23.0$^{\prime\prime}$ & 0.46 & 3.69/3.95 & $259\pm 17$ & +$0.65\pm 0.08$\\ \hline
J15541 & J155413.0--560932 & 15h54m13.09s & --56$^{\circ}$09$^{\prime}$32.6$^{\prime\prime}$ & 0.75 & 3.54/4.32 & $44\pm 8$ & +$0.60\pm 0.21$\\ \hline
J16181 & J161807.7--540612 & 16h18m07.74s & --54$^{\circ}$06$^{\prime}$12.5$^{\prime\prime}$ & 0.50 & 0.04/4.61 & $253\pm 17$ & +$0.94\pm 0.09$\\ \hline
J16246a & J162425.2--460316 & 16h24m25.20s & --46$^{\circ}$03$^{\prime}$16.5$^{\prime\prime}$ & 0.67 & 7.92/4.61 & $660\pm 27$ & +$0.14\pm 0.04$\\
J16246b & J162430.7--455514 & 16h24m30.77s & --45$^{\circ}$55$^{\prime}$14.1$^{\prime\prime}$ & 0.46 & 1.56/4.61 & $125\pm 12$ & +$0.62\pm 0.12$\\ \hline
J17096 & J170950.2--252934 & 17h09m50.27s & --25$^{\circ}$29$^{\prime}$34.4$^{\prime\prime}$ & 0.55 & 2.02/2.73 & $34\pm 7$ & +$0.12\pm 0.22$\\ \hline
J17118 & J171135.8--315504 & 17h11m35.89s & --31$^{\circ}$55$^{\prime}$04.5$^{\prime\prime}$ & 0.81 & 3.08/2.84 & $23\pm 6$ & +$0.30\pm 0.29$\\ \hline
J17508a & J175106.8--321827 & 17h51m06.85s & --32$^{\circ}$18$^{\prime}$27.9$^{\prime\prime}$ & 0.44 & 3.21/2.31 & $713\pm 28$ & --$0.55\pm 0.04$\\
J17508b & J175108.7--322122 & 17h51m08.78s & --32$^{\circ}$21$^{\prime}$22.2$^{\prime\prime}$ & 0.48 & 3.68/2.31 & $169\pm 14$ & +$0.03\pm 0.09$\\ \hline
J18007 & J180042.6--414650 & 18h00m42.69s & --41$^{\circ}$46$^{\prime}$50.0$^{\prime\prime}$ & 0.51 & 1.68/3.19 & $441\pm 22$ & +$0.27\pm 0.05$\\ \hline
J19294 & J192930.1+132705 & 19h29m30.14s & +13$^{\circ}$27$^{\prime}$05.7$^{\prime\prime}$ & 0.39 & 0.10/3.43 & $39\pm 7$ & +$1.00\pm 0.27$\\ \hline
J20310 & J203055.2+383347 & 20h30m55.28s & +38$^{\circ}$33$^{\prime}$47.1$^{\prime\prime}$ & 0.35 & 1.39/4.54  & $266\pm 17$ & +$0.86\pm 0.09$\\ \hline
\end{tabular}
\end{minipage}
\end{table}

\begin{table}
\caption{{\em Chandra} Spectral Parameters and Absolute Spurious Probabilities \label{tab:spectra_ch}}
\begin{minipage}{\linewidth}
\footnotesize
\begin{tabular}{ccccccccc} \hline \hline
IGR Name & $N_{\rm H}$\footnote{The errors on the parameters are 90\% confidence. The column density is calculated assuming \cite{wam00} abundances and \cite{vern96} cross sections.} & $N_{\rm H,Galactic}$\footnote{From the HI4PI survey \citep{bb16}.} & $\Gamma$ & Absorbed Flux\footnote{In units of erg\,cm$^{-2}$\,s$^{-1}$.} & Unabsorbed Flux\footnote{Only corrected for Galactic absorption.} & $C$/dof & $P_{\rm reject}$\footnote{The probability that an absorbed power-law does not provide a good description of the spectrum based on a calculation of the variance of $C$ according to the method described in \cite{kaastra17}.} & Probability\footnote{Absolute probability that a source of this brightness would be found by chance in the search region.}\\
  &  ($\times$$10^{22}$\,cm$^{-2}$) & ($\times$$10^{22}$\,cm$^{-2}$) & & (2--10\,keV) & (2--10\,keV) & & & \\ \hline \hline
J15038 & $<$1.0 & 1.3 & $0.0\pm 0.3$ & $2.34\times 10^{-12}$ & $2.35\times 10^{-12}$ & 25/23 & 21\% & 0.97--1.03\%\\
J15541 & $<$4.3 & 0.8 & $1.1\pm 1.1$ & $3.54\times 10^{-13}$ & $3.78\times 10^{-13}$ & 5.8/6 & -- & 5.4--10.7\%\\
J16181 & $9\pm 7$ & 0.5 & $0.8\pm 1.1$ & $1.27\times 10^{-12}$ & $1.29\times 10^{-12}$ & 5.0/7 & -- & 2.3--2.8\%\\
J16246a & $0.4\pm 0.3$ & 0.8 & $1.5\pm 0.2$ & $2.82\times 10^{-12}$ & $2.89\times 10^{-12}$ & 90/57 & 99.7\%  & 3.0--3.5\%\\
J16246b & $2.5\pm 1.6$ & 0.8 & $1.3\pm 0.7$ & $9.27\times 10^{-13}$ & $1.04\times 10^{-12}$ & 16/10 & 75\%  & 2.9--3.5\%\\
J17096 & $<$3.6 & 0.2 & $1.9\pm 1.3$ & $1.33\times 10^{-13}$ & $1.45\times 10^{-13}$ & 5.6/4 & 31\% & 4.7--13.7\%\\
J17118 & $<$4.2 & 0.3 & $1.2^{+1.7}_{-1.1}$ & $3.22\times 10^{-13}$ & $3.37\times 10^{-13}$ & 2.8/2 & 19\% & 3.9--6.4\%\\
J17508a & $0.2\pm 0.1$ & 0.5 & $3.0\pm 0.2$ & $6.22\times 10^{-13}$ & $6.33\times 10^{-13}$ & 75/59 & 84\%  & 1.9--3.2\%\\
J17508b & $0.5\pm 0.5$ & 0.5 & $1.8\pm 0.5$ & $4.87\times 10^{-13}$ & $5.00\times 10^{-13}$ & 17/13 & 51\% & 3.1--5.6\%\\
J18007 & $<$0.14 & 0.12 & $0.8\pm 0.2$ & $1.06\times 10^{-12}$ & $1.06\times 10^{-12}$ & 23/14 & 88\%  & 1.3--1.7\%\\
J19294 & $28^{+38}_{-24}$ & 0.8 & $0.6^{+2.7}_{-2.2}$ & $5.84\times 10^{-13}$ & $5.91\times 10^{-13}$ & 2.1/5 & -- & 2.3--4.0\%\\
J20310 & $<$5.6 & 1.5 & --$0.2\pm 0.9$ & $1.95\times 10^{-12}$ & $2.04\times 10^{-12}$ & 9.2/9 & 3\% & 1.5--1.6\%\\ \hline
\end{tabular}
\end{minipage}
\end{table}

\begin{table}
\caption{{\em Chandra}+{\em INTEGRAL} Spectral Parameters\label{tab:spectra}}
\begin{minipage}{\linewidth}
\footnotesize
\begin{tabular}{cccccc} \hline \hline
IGR Name & $N_{\rm H}$\footnote{The errors on the parameters are 90\% confidence. The column density is calculated assuming \cite{wam00} abundances and \cite{vern96} cross sections.} & $\Gamma$ & $E_{\rm fold}$ & Unabsorbed Flux\footnote{In units of erg\,cm$^{-2}$\,s$^{-1}$, and corrected for Galactic absorption.} & $\chi^{2}$/dof\\
  & ($\times$$10^{22}$\,cm$^{-2}$) & & (keV) & (0.3--100\,keV) & \\ \hline\hline
J15038 & $<$1.0 & --$0.3^{+0.6}_{-0.3}$ & $10^{+8}_{-2}$ & $1.11\times 10^{-11}$ & 22/24\\
J15541 & $<$2.4 & $0.6^{+0.2}_{-0.6}$ & $>$36 & $1.07\times 10^{-11}$ & 5/7\\
J16181 & $8\pm 5$ & $0.3\pm 0.7$ & $20^{+33}_{-8}$ & $1.10\times 10^{-11}$ & 5/8\\
J16246a & $0.3\pm 0.3$ & $1.4\pm 0.3$ & $>$22 & $1.01\times 10^{-11}$ & 86/58\\
J16246b & $1.4\pm 1.3$ & $0.6\pm 0.7$ & $>$12 & $7.6\times 10^{-12}$ & 18/11\\
J17096\footnote{The quality of the fit does not allow for errors to be calculated.  The {\ttfamily tbabs*cutoffpl} does not provide a good description of the spectrum.} & 0.0 & --2.0 & 9 & $1.15\times 10^{-11}$ & 21/5\\
J17118 & $<$1.4 & $0.2^{+1.1}_{-0.7}$ & $>$8 & $4.2\times 10^{-12}$ & 6/3\\
J17508a$^{c}$ & 0.15 & 2.9 & 500 & $2.2\times 10^{-12}$\footnote{This is the flux for the model shown in Figure~\ref{fig:spectra}, which is well below the {\em INTEGRAL} measurements.} & 192/60\\
J17508b & $<$0.8 & $0.8\pm 0.1$ & $>$179 & $1.35\times 10^{-11}$ & 43/14\\
J18007 & $<$0.12 & $0.1\pm 0.3$ & $25^{+13}_{-8}$ & $1.67\times 10^{-11}$ & 43/15\\
J19294 & $24^{+21}_{-14}$ & --$0.1\pm 0.9$ & $20^{+42}_{-8}$ & $1.08\times 10^{-11}$ & 2/6\\
J20310 & $<$4.6 & --$0.9\pm 1.1$ & $7^{+10}_{-3}$ & $1.14\times 10^{-11}$ & 9/10\\ \hline
\end{tabular}
\end{minipage}
\end{table}

\begin{table}
\caption{{\em Gaia} Identifications\label{tab:gaia}}
\begin{minipage}{\linewidth}
\footnotesize
\begin{tabular}{cccccc} \hline \hline
IGR Name & {\em Gaia} Number & Separation\footnote{The angular separation between the {\em Chandra} position and the {\em Gaia} catalog position.} & $G$-Magnitude & Parallax & Distance\footnote{From \cite{bj18}.}\\ 
 & (in DR2) & (arcsec) &  & (milliarcsec) & (kpc)\\\hline \hline
J15038  & 5876459780108921216 & 0.072 & $19.055\pm 0.006$ &   $1.09\pm 0.37$ & $1.08^{+1.53}_{-0.42}$\\
J15541  & 5836092447721411584 & 0.245 & $20.392\pm 0.019$ & --$6.92\pm 1.87$ & ---\\
J16246a & 5942431027538084608 & 0.463 & $15.494\pm 0.003$ &   $0.43\pm 0.05$ & $2.20^{+0.28}_{-0.22}$\\
J17096  & 4112378173142821248 & 0.049 & $19.360\pm 0.006$ & --$0.52\pm 0.43$ & ---\\
J17508a & 4043524762244704000 & 0.240 & $9.027\pm 0.001$  &   $7.33\pm 0.04$ & $0.135^{+0.002}_{-0.001}$\\
J17508b & 4043518508770139648 & 0.383 & $18.884\pm 0.018$ &   ---            & ---\\
J18007  & 6725376279628784384 & 0.331 & $16.009\pm 0.006$ &   $0.38\pm 0.06$ & $2.49^{+0.49}_{-0.36}$\\ \hline
\end{tabular}
\end{minipage}
\end{table}

\begin{table}
\caption{{\em WISE} Identifications\label{tab:wise}}
\begin{minipage}{\linewidth}
\footnotesize
\begin{tabular}{cccccccl} \hline \hline
IGR Name & AllWISE Name & Separation\footnote{The angular separation between the {\em Chandra} position and the AllWISE catalog position.} & $W1$ & $W2$ & $W3$ & $W4$ & ex\footnote{A ``0'' indicates that the source is consistent with being a point source, and a ``1'' indicates that the profile is not well-described as a point source in at least one photometric band.} \\ 
         &              & (arcsec)        &  \\\hline \hline
J16181  & J161807.75--540612.3 & 0.21 & $11.29\pm 0.04$ & $10.12\pm 0.02$ & $7.55\pm 0.02$ & $4.97\pm 0.04$ & 1\\
J16246a & J162425.20--460316.7 & 0.28 & $11.35\pm 0.03$ & $11.37\pm 0.02$ & $11.88\pm 0.38$ & --- & 0\\
J16246b & J162430.78--455514.4 & 0.35 & $10.70\pm 0.02$ & $10.52\pm 0.02$ & $8.74\pm 0.04$ & $6.58\pm 0.07$ & 1\\
J17096  & J170950.25--252934.7 & 0.41 & $13.61\pm 0.04$ & $12.54\pm 0.03$ & $9.40\pm 0.06$ & $6.78\pm 0.14$ & 1\\
J17508a & J175106.84--321827.8 & 0.11 &  $7.41\pm 0.03$ & $7.48\pm 0.02$ & $7.29\pm 0.02$ & $6.52\pm 0.09$ & 0\\
J19294  & J192930.14+132705.9  & 0.32 & $12.56\pm 0.03$ & $12.28\pm 0.03$ & $9.71\pm 0.05$ & $7.38\pm 0.12$ & 1\\ \hline
\end{tabular}
\end{minipage}
\end{table}

\begin{table}
\caption{VISTA, 2MASS, and UKIDSS Identifications\label{tab:nir}}
\begin{minipage}{\linewidth}
\footnotesize
\begin{tabular}{cccccccc} \hline \hline
IGR Name & Catalog & Source (arcsec) & Separation\footnote{The angular separation between the {\em Chandra} position and the catalog position.} & $J$ & $H$ & $K$/$K_{s}$ & Class\footnote{The classification based on the spatial profile, where --2 is a probable star, --1 is a star with probability $\ge$90\%, and 1 is a galaxy with probability $\ge$90\%.}\\\hline \hline
J15038  & VISTA & VVV J150415.72--602122.87 & 0.230 & $16.13\pm 0.02$ & $15.48\pm 0.02$ & $K_{s}=15.03\pm 0.02$ & --1\\
J15541  & --- & --- & --- & --- & --- & See Figure~\ref{fig:nir}.& ---\\
J16181  & 2MASS & 2MASS J16180771--5406122 & 0.365 & --- & $14.09\pm 0.04$ & $K_{s}=12.85\pm 0.06$ & ---\\
J16246a & 2MASS & 2MASS J16242520--4603169 & 0.409 & $12.61\pm 0.02$ & $11.78\pm 0.02$ & $K_{s}=11.49\pm 0.02$ & ---\\
J16246b & 2MASS & 2MASS J16243080--4555144 & 0.519 & $14.31\pm 0.10$ & $12.98\pm 0.11$ & $K_{s}=12.19\pm 0.08$ & ---\\
J17096  & VISTA & VHS 472814537075   & 0.135 & $17.06$ & --- & $K_{s}=15.81$ & --1\\
J17118  & VISTA & VVV J171135.91--315503.61 & 0.929 & $18.64\pm 0.28$ & $18.00\pm 0.33$ & $K_{s}=16.97\pm 0.20$ & --2\\
J17508a & 2MASS & 2MASS J17510684--3218276 & 0.054 & $7.89\pm 0.02$ & $7.59\pm 0.04$ & $K_{s}=7.45\pm 0.03$ & ---\\
J17508b & VISTA  & VVV J175108.76--322122.39 & 0.306 & $14.73\pm 0.02$ & $13.63\pm 0.02$ & $K_{s}=13.17\pm 0.01$ & --1\\
J18007  & VISTA  & VVV J180042.71--414650.23 & 0.339 & $15.60\pm 0.01$ & $15.34\pm 0.01$ & $K_{s}=15.31\pm 0.03$ & --1\\
J19294  & UKIDSS & UGPS J192930.11+132705.7 & 0.372 & $16.30\pm 0.01$ & $15.09\pm 0.01$ & $K=14.32\pm 0.01$ & 1\\
J20310  & UKIDSS & UGPS J203055.29+383347.1 & 0.180 & $18.83\pm 0.05$ & $17.30\pm 0.02$ & $K=16.54\pm 0.04$ & 1\\ \hline
\end{tabular}
\end{minipage}
\end{table}

\begin{table}
\caption{Summary of Results\label{tab:summary}}
\begin{minipage}{\linewidth}
\footnotesize
\begin{tabular}{cccl} \hline \hline
IGR Name & {\em Chandra} counterpart or & Source Type & Evidence\\
         & 2--10\,keV flux limit  &             &         \\ \hline
J03599+5043  & $<$$1.5\times 10^{-13}$ & --- & ---\\
J07202+0009  & $<$$1.4\times 10^{-13}$ & --- & ---\\
J15038--6021 & CXOU J150415.7--602123 & CV/IP? & {\em Gaia} distance, near-IR magnitudes, X-ray spectrum\\
J15541--5613 & CXOU J155413.0--560932 & AGN? & X-ray spectrum\\
J16120--3543 & CXOU J161147.0--354634? & AGN? & {\em WISE} colors\\
J16181--5407 & CXOU J161807.7--540612 & AGN & {\em WISE} colors and extended in near-IR\\
J16246--4556 & CXOU J162430.7--455514 & AGN & Extended in near-IR\\
J16482--2959 & $<$$1.5\times 10^{-13}$ & --- & ---\\
J17096--2527 & CXOU J170950.2--252934 & AGN & {\em WISE} colors and extended in near-IR\\
J17118--3155 & CXOU J171135.8--315504 & ? & ?\\
J17508--3219 & CXOU J175108.7--322122? & ?\footnote{While CXOU J175108.7--322122 is classified as a Dwarf Nova and may be the counterpart, source \#1 detected in the field by {\em Swift} \citep{landi17} is a variable source that is another possible counterpart.} & ?\\
J18007--4146 & CXOU J180042.6--414650 & CV/IP? & {\em Gaia} distance, near-IR magnitudes, X-ray spectrum\\
J19294+1327  & CXOU J192930.1+132705 & AGN & Extended in near-IR\\
J20310+3835  & CXOU J203055.2+383347 & AGN? & Likely extended in near-IR\\
J20413+3210  & $<$$1.5\times 10^{-13}$ & --- & ---\\ \hline\hline
\end{tabular}
\end{minipage}
\end{table}


\clearpage

\end{document}